\documentstyle[amstex,aps]{revtex}

\begin{document}
\draft
\title{A Geometric Approach to the Standard Model}
\author{Greg Trayling}
\address{Department of Physics, University of Windsor, Windsor, Ontario,\\
Canada N9B 3P4}
\date{November 29, 1999}
\maketitle

\begin{abstract}
A geometric approach to the standard model in terms of the Clifford algebra $%
C\!\ell _{7}$ is advanced. The gauge symmetries and charge assignments of
the fundamental fermions are seen to arise from a simple geometric model
involving extra space-like dimensions. The bare coupling constants are found
to obey $g_{s}/g=1$ and $g^{\prime }/g=\sqrt{3/5}$, consistent with $SU(5)$
grand unification but without invoking the notion of master groups. In
constructing the Lagrangian density terms, it is found that the Higgs
isodoublet field emerges in a natural manner. A matrix representation of $%
C\!\ell _{7}$ is included as a computational aid.
\end{abstract}

\pacs{PACS number(s): 12.10.Dm, 11.10.Kk, 12.60.-i}


\section{Introduction}

In many physics equations of a fundamental nature, Clifford (geometric)
algebras\cite{Lounesto} may be employed to recast conventional expressions
into more holistic and aesthetically pleasing forms. This approach often
reveals insights which were previously obscured by an inappropriate choice
of architecture. The result may allow the consolidation of incongruous
terms, suggest missing pieces, or reveal new restrictions imposed by
identifying privileged subspaces within the chosen algebra.

The present paper explores the minimal standard model in terms of the
Clifford algebra $C\!\ell _{7}$. The aim here is to demonstrate that the
seemingly disparate gauge symmetries $U(1)_{Y}\otimes SU(2)_{L}\otimes
SU(3)_{C}$ may be unified under a single simple geometric model. The implied
formalism is then used to construct analogues of the conventional Lagrangian
density terms.

In Sec. II the basic algebraic operations and conventions are laid out, as
much of the notation used varies considerably within the Clifford algebra
community. A brief recapitulation of the application of $C\!\ell _{3}$ to
the description of flat space-time, as developed by Baylis\cite{Baylis1}, is
presented as the proper foundation for the addition of higher space-like
dimensions. A matrix representation of $C\!\ell _{7}$ is offered in order to
better elucidate the structure of the higher-dimensional spinors used and to
demonstrate the connection to the conventional Dirac algebra. Section III
details the gauge symmetries of the standard model as plane-rotational
invariances of the $C\!\ell _{3}$ (physical) part of an extended
current-density expression involving all of the fermions of a single
generation. The various group generators are worked out explicitly for the
specific spinor representation used. In Sec. IV the gauge formalism
developed is then applied to the construction of each of the various terms
in the Lagrangian density. The bare coupling constants then follow
immediately from the normalization of the double-sided set of generators. It
is also shown that the minimal Higgs field\cite{Higgs} has a natural origin,
involving higher-dimensional components of the current.

\section{Algebraic Foundations}

In the real Clifford algebra $C\!\ell _{3}$, the vector elements $%
\{e_{1},e_{2},e_{3}\}$ are chosen to represent the three physical space-like
directions. The product of any number of vectors is completely determined by
the anticommutator
\begin{equation}
\{e_{j},e_{k}\}=2\delta _{jk}.  \label{commutate}
\end{equation}
All higher-order products of the vectors can be reduced to linear
combinations of the 8 basis forms $%
\{1;e_{1},e_{2},e_{3};e_{1}e_{2},e_{2}e_{3},e_{3}e_{1};e_{1}e_{2}e_{3}\}$.
The three bivector forms and single trivector form are taken to represent
planes and the volume element respectively\cite{Baylis2}.

Two basic real conjugations which are anti-automorphic involutions are
frequently used in this paper. The reversion of $K\in C\!\ell _{3}$, denoted
$K^{\dagger }$, is obtained by reversing the order of appearance of all
vector elements within $K$. For example,
\begin{equation}
(e_{1}e_{2}e_{3})^{\dagger }=e_{3}e_{2}e_{1}=-e_{1}e_{2}e_{3}.  \label{ex1}
\end{equation}
Both reversing the order and negating all vector elements of $K$ defines
Clifford conjugation, denoted by $\bar{K}$. For example,
\begin{equation}
\overline{(e_{1}e_{2})}=(-e_{2})(-e_{1})=-e_{1}e_{2}.  \label{ex2}
\end{equation}
Within $C\!\ell _{3}$, the latter operation serves to negate the vector and
bivector portions while leaving the scalar and trivector invariant. For both
of these operations we clearly have
\begin{eqnarray}
\overline{(AB)} &=&\bar{B}\bar{A},  \nonumber \\
(AB)^{\dagger } &=&B^{\dagger }A^{\dagger }.  \label{product}
\end{eqnarray}

One type of structure that will be of central importance in this paper is
that of projectors---pairs of real idempotent terms $P_{\pm }$ satisfying
\begin{eqnarray}
P_{\pm }^{2}=P_{\pm } &=&P_{\pm }^{\dagger },  \nonumber \\
P_{\pm }P_{\mp } &=&0,  \nonumber \\
P_{+}+P_{-} &=&1.
\end{eqnarray}
An example in $C\!\ell _{3}$ are the elements
\begin{equation}
P_{\pm 3}=\frac{1}{2}(1\pm e_{3}).
\end{equation}

In applying this formalism to relativistic physics, time may be allocated to
the scalar part of a general real vector $V=v^{0}+v^{j}e_{j}$. The Minkowski
metric then arises naturally through the norm of $V$ given by
\begin{equation}
V\bar{V}=\lceil v^{\mu }v_{\mu }\rfloor .  \label{norm}
\end{equation}
The delimiters in Eq.\ (\ref{norm}) will be used throughout this paper to
designate the prevailing non-algebraic notation, outside of which there is
no metric implied in the summation convention for repeated indices.

One of the agreeable aspects of this formalism is that the Lorentz
transformations have a particularly lucid form. Proper and orthochronous
Lorentz transformations of vectors are represented by
\begin{equation}
V\rightarrow LVL^{\dagger },  \label{Lorentz}
\end{equation}
where $L$ is any unimodular element: $L\bar{L}=1,$ which can be expressed as
the product\cite{Baylis3}
\begin{equation}
L=\exp (\eta /2)\exp (-\theta /2),  \label{L}
\end{equation}
where $\eta =\eta ^{i}e_{i}$ and $\theta =\theta ^{i}\epsilon
_{ijk}e_{j}e_{k}$. Pure rotations $L_{R}=\exp (-\theta /2{\bf )}$ are
characterized by the unit plane of rotation $\epsilon
_{ijk}e_{j}e_{k}/(-\theta \overline{\theta })^{1/2}$. Similarly, pure boosts
$L_{B}=\exp (\eta /2)$ are specified by the rapidity $(-\eta \bar{\eta}%
)^{1/2}$ directed along a unit vector in the direction of the boost. The
clear geometric interpretation of this method will be employed later when
the gauge transformations are introduced.

Objects such as Eq.\ (\ref{norm}) are clearly Lorentz invariant scalars,
since by Eqs.\ (\ref{product}) and (\ref{Lorentz}) we have
\begin{equation}
V\bar{V}\rightarrow LV(L^{\dagger }\bar{L}^{\dagger })\bar{V}\bar{L}=LV\bar{V%
}\bar{L}=(L\bar{L})V\bar{V}=V\bar{V}.
\end{equation}
One must be careful here to distinguish between invariant scalars and the
time component of a four-vector, which both occupy the same place in the
algebra but transform differently through Eq.\ (\ref{Lorentz}). The notation
$e_{0}$ will often be used for the time component, with the understanding
that it is algebraically just the identity element satisfying $\bar{e}%
_{0}=e_{0}$. For example, a Lorentz invariant trivector representing the
full space-time volume element can be written as
\begin{equation}
e_{0}\bar{e}_{1}e_{2}\bar{e}_{3}\rightarrow L(e_{0}\bar{e}_{1}e_{2}\bar{e}%
_{3})\bar{L}=e_{0}\bar{e}_{1}e_{2}\bar{e}_{3}.
\end{equation}
Again, we make no algebraic distinction between this and the trivector $%
e_{1}e_{2}e_{3},$ leaving it to be implicitly understood by the
transformation rules which one is meant. In general, the same algebra will
be used to construct a number of different objects, which may be further
classified by how they transform.

Spinors may be defined as entities which transform according to the rule
\begin{equation}
\psi \rightarrow L\psi .  \label{spinor}
\end{equation}
Under this prescription, spinors span all of the basis forms, but contain
some projector structure on the opposite side of the Lorentz transformation
operator. To illustrate this point in $C\!\ell _{3}$, consider the spinor
\begin{equation}
\psi =\sqrt{2}(\phi _{0r}+\phi _{1r}e_{1}+\phi _{1i}e_{2}e_{3}+\phi
_{0i}e_{1}e_{2}e_{3})P_{+3},  \label{rspinor}
\end{equation}
which has been contrived to transform under (\ref{spinor}) in the same
manner as the right-chiral\footnote{%
To avoid confusion with later double-sided transformations, the term
``-chiral'' is preferred in place of ``-handed''.} portion of the
conventional Weyl representation column spinor\cite{Weyl}. Once $\psi $ is
explicitly defined in this manner, the projector structure is unaltered by
any Lorentz transformation. Conversely, multiplying any fixed element onto $%
\psi $ from the right will not affect the Lorentz transformations. Writing a
current vector as
\begin{equation}
\psi \psi ^{\dagger }=\lceil \bar{\psi}\gamma ^{\mu }\psi \rfloor e_{\mu },
\end{equation}
we can see that the choice of projector in this case is not unique, since $%
\psi $ may be multiplied on the right by a rotation operator $R$, having the
same form as $L_{R}$. This leaves the current invariant by
\begin{equation}
\psi \psi ^{\dagger }\rightarrow \psi RR^{\dagger }\psi ^{\dagger }=\psi R%
\bar{R}\psi ^{\dagger }=\psi \psi ^{\dagger },
\end{equation}
and rotates the spatial vector in the projector through
\begin{equation}
\psi (1+e_{3})R_{\theta }=\psi R_{\theta }(1+R_{-\theta }e_{3}R_{-\theta
}^{\dagger }).
\end{equation}
One might be tempted to use this as the basis for some SU(2) current
symmetry, but instead we shall fix the spatial projector into a specific
direction by defining the current as
\begin{equation}
J=\psi P_{+3}\psi ^{\dagger }.  \label{current}
\end{equation}
Since $P_{+3}^{\dagger }=P_{+3}$, we need now only consider the $P_{+3}$
spinor forms. The current is then viewed as a sort of `pre-Lorentz'
transformation of the light-like vector $\frac{1}{2}(e_{0}+e_{3})$, with the
spinors acting as the transformation operation. The Lorentz transformation
of the current is then
\begin{equation}
J\rightarrow L\psi P_{+3}\psi ^{\dagger }L^{\dagger }=LJL^{\dagger },
\end{equation}
where the internal structure reflects a choice in the form of the spinors
and is treated as a Lorentz invariant. The form of Eq.\ (\ref{current})
still admits a geometric U(1) gauge symmetry, since an internal rotation of
the projector about the plane $e_{1}e_{2}$ yields
\begin{equation}
\psi _{+3}\exp (\theta ^{3}e_{1}e_{2}/2)=\exp (i\theta ^{3}/2)\psi _{+3}.
\end{equation}
This seemingly innocuous restriction that the physical projector should
remain gauge invariant will now be extended to a higher-dimensional space,
where a more elaborate projector structure on the spinors leads to an
interesting framework in which the complete gauge symmetry of the standard
model arises in a natural manner.

\subsection{Higher Dimensions}

Each extra spatial dimension beyond the three physical vectors $%
\{e_{1},e_{2},e_{3}\}$ is accommodated simply by introducing a new $e_{j}$
in accordance with Eq.\ (\ref{commutate}). The commutation of any
higher-dimensional $(j>3)$ vector with any element $K\in C\!\ell _{3}$ may
be summarized by
\begin{equation}
e_{j}K=\bar{K}^{\dagger }e_{j}.  \label{hi-com}
\end{equation}
The definitions of reversal $(K^{\dagger })$ and Clifford conjugation $(\bar{%
K})$ exemplified in Eqs.\ (\ref{ex1}) and (\ref{ex2}) are extended to
encompass higher dimensions with no additional modifications. As vector
components, each of these extra dimensions is then manifestly Lorentz
invariant. For example,
\begin{equation}
e_{4}\rightarrow Le_{4}L^{\dagger }=L\bar{L}e_{4}=e_{4}.
\end{equation}
No further presumptions are made regarding the true nature of these
additional dimensions, other than that they are, in a sense, inaccessible
directly in flat space-time if Lorentz transformations cannot betray their
presence.

Rotations in higher dimensions, which are key to the forthcoming discussion
of gauge symmetries, are managed by directly extending the form of Eq.\ (\ref
{L}) to accommodate the exponentiation any element proportional to a unit
bivector in $C\!\ell _{7}$ representing the plane of rotation. Since
rotations in $C\!\ell _{3}$ leave all higher dimensions invariant, it is
trivial to map the same framework to any plane in $C\!\ell _{7}$ by
symmetry. Note that in euclidean spaces of dimensionality greater than
three, there are an infinite number of planes orthogonal to a given
direction. It is no longer sufficient outside of $C\!\ell _{3}$ to
characterize a rotation by an axis.

The choice of adding four extra space-like dimensions $%
\{e_{4},e_{5},e_{6},e_{7}\}$ to form $C\!\ell _{7}$ is arrived at by simple
counting arguments. Assuming four complex numbers (eight basis forms) for
each fermion, one generation consisting of an electron, a neutrino (assuming
both left- and right-chiral), three up quarks and three down quarks,
requires 64 basis forms for the spinors. Each additional dimension doubles
the number of basis forms, so $C\!\ell _{7}$, with 128 basis forms, is
appropriate since only half of these will be used under the $P_{+3}$
projection restriction.

$C\!\ell _{7}$ is also attractive in that the total volume element of the
algebra
\begin{equation}
e_{t}\equiv e_{1}e_{2}e_{3}e_{4}e_{5}e_{6}e_{7}
\end{equation}
commutes with all of the basis forms and squares to $-1$, and can therefore
be identified with the unit imaginary. This occurs for every $C\!\ell
_{4n+3} $. If one were to settle for $C\!\ell _{6}$, for example, the full
volume element of that algebra would anticommute with all vectors, and one
would be compelled to consider its inclusion as an additional `dimension'
regardless. This is also the case with the Dirac algebra, since $\gamma
^{5}\equiv i\gamma ^{0}\gamma ^{1}\gamma ^{2}\gamma ^{3}$ anticommutes with
all $\gamma ^{\mu }$.

\subsection{$C\!\ell _{7}$ Matrix Representation}

Building upon the Dirac algebra defined by
\begin{equation}
\{\gamma ^{\mu },\gamma ^{\nu }\}=2g^{\mu \upsilon },  \label{Dirac}
\end{equation}
a faithful $8\times 8$ matrix representation of $C\!\ell _{7}$ may be
constructed by
\begin{eqnarray}
1 &\sim &\left(
\begin{array}{ll}
1 & 0 \\
0 & 1
\end{array}
\right) ,\quad e_{k}\sim \left(
\begin{array}{cc}
\gamma ^{0}\gamma ^{k} & 0 \\
0 & -\gamma ^{0}\gamma ^{k}
\end{array}
\right) ,\quad e_{4}\sim \left(
\begin{array}{cc}
\gamma ^{1}\gamma ^{2}\gamma ^{3} & 0 \\
0 & -\gamma ^{1}\gamma ^{2}\gamma ^{3}
\end{array}
\right) ,  \nonumber \\
e_{5} &\sim &\left(
\begin{array}{cc}
\gamma ^{0} & 0 \\
0 & -\gamma ^{0}
\end{array}
\right) ,\quad e_{6}\sim \left(
\begin{array}{cc}
0 & 1 \\
1 & 0
\end{array}
\right) ,\quad e_{7}\sim \left(
\begin{array}{cc}
0 & i \\
-i & 0
\end{array}
\right) .  \label{matrices}
\end{eqnarray}
It is trivial to prove by contradiction that since the equivalent matrices
satisfy the anticommutator (\ref{commutate}), the matrix representations of
the 128 basis forms over the reals span all complex $8\times 8$ matrices.
The total volume element $e_{t}$ is equivalent to the diagonal matrix $%
[i]_{8\times 8}$ and will be used interchangeably with the unit imaginary.
For example, we may write algebraically
\begin{equation}
e_{1}e_{2}e_{3}=ie_{4}e_{5}e_{6}e_{7}.
\end{equation}
This affords a complex space which will be useful when comparing algebraic
terms to those formulated in conventional notation. Note that of the 128
basis forms, only $1$ and $i$ have traceless matrix representations,
therefore the complex scalar portion of any algebraic term $K\in C\!\ell
_{7} $ may be extracted by calculating the trace $\frac{1}{8}Tr(M)$ of its
equivalent matrix. Embedding the Dirac algebra in this manner supplants the
need to introduce any explicit artificial metric as in Eq.\ (\ref{Dirac}).
This is advanced as the proper means to add extra space-like dimensions to
the Dirac algebra.

In developing the structure of spinors, the columns of an arbitrary complex $%
8\times 8$ matrix are factored out in terms of the $C\!\ell _{7}$ basis
forms. This is directly analogous to representing spinors as column matrices
in the conventional Dirac formalism, except now we may include several
distinct fermion fields in the same term by using different columns for
each. It is illustrative to adopt a particular $\gamma $ matrix
representation so that the columns may be factored explicitly. The Weyl
representation is again chosen here because it yields a clear partition
between the right- and left-chiral spinor components in both the column and
algebraic forms.

Consider the top four complex elements of the first column, which are
equivalent to

\begin{equation}
\left(
\begin{array}{l}
\phi _{0} \\
\phi _{1} \\
\phi _{2} \\
\phi _{3}
\end{array}
\right) \sim [(\phi _{0}+\phi _{1}e_{1})-(\phi _{2}-\phi
_{3}e_{1})e_{5}]P_{+3}P_{-\alpha }P_{+\beta }.
\end{equation}
This retains a core spinor structure similar to that of spinors in $C\!\ell
_{3}$, multiplied by two new projectors which have been abbreviated by
\begin{eqnarray}
P_{\pm \alpha } &\equiv &\frac{1}{2}(1\pm ie_{4}e_{5}),  \nonumber \\
P_{\pm \beta } &\equiv &\frac{1}{2}(1\pm ie_{6}e_{7}).
\end{eqnarray}
The bottom four components of the first column translate as
\begin{equation}
\left(
\begin{array}{l}
\zeta _{2} \\
\zeta _{3} \\
\zeta _{0} \\
\zeta _{1}
\end{array}
\right) \sim [(\zeta _{0}+\zeta _{1}e_{1})e_{5}e_{6}+(\zeta _{2}-\zeta
_{3}e_{1})e_{6}]P_{+3}P_{-\alpha }P_{+\beta },
\end{equation}
which has the same projector structure as the upper four components. The
chiral designation has been interchanged in accordance with the conjugation
of the lower $\gamma ^{0}\gamma ^{k}$ submatrix in $e_{k}$. Each of the
eight possible products $P_{\pm 3}P_{\pm \alpha }P_{\pm \beta }$ corresponds
to a real diagonal matrix which projects out the column associated with each
projector set. For example,
\begin{equation}
P_{1}\equiv P_{+3}P_{-\alpha }P_{+\beta }\sim diag(1,0,...,0).
\end{equation}

Anticipating that an SU(2) doublet will arise from the degrees of freedom
within a given column, and considering the submatrix structure in (\ref
{matrices}), the upper and lower four components are ascribed to distinct
fermions. Transcribing the remaining columns into algebraic form using the
Weyl representation, the four columns corresponding to $P_{+3}$ projections
are summarized in table \ref{table1}, where
\begin{eqnarray}
\psi _{R} &\equiv &(\phi _{0}+\phi _{1}e_{1})P_{+3},  \nonumber \\
\psi _{L} &\equiv &(\phi _{3}-\phi _{2}e_{1})P_{+3},  \label{Weyl}
\end{eqnarray}
and the factor of $\sqrt{8}$ has been inserted for later normalizations. In
general, other representations have a similar projector structure and are
accessible through a different choice in Eq.\ (\ref{Weyl}).

Evidently, the chiral projector for all of the fermions in table \ref{table1}
is given by
\begin{equation}
P_{R/L}=\frac{1}{2}(1\pm e_{4}e_{5}e_{6}e_{7})
\end{equation}
operating from the left-hand side. This is easily seen by monitoring the
sign changes in the element
\begin{equation}
e_{4}e_{5}e_{6}e_{7}=-(ie_{4}e_{5})(ie_{6}e_{7})
\end{equation}
as it both passes through the core spinor, where it commutes with any
physical vector, anticommutes with any higher-dimensional vector, and is
then absorbed into the column projectors. The parity operator is given by
\begin{equation}
P:\Psi \rightarrow \Psi ^{\prime }=e_{1}e_{2}e_{3}e_{4}\Psi ,  \label{parity}
\end{equation}
where $\Psi $ denotes the entire spinor set of table \ref{table1}. Both the
projector and operator also have clear matrix representations through (\ref
{matrices}).

The projectors $\{P_{\pm 3},P_{\pm \alpha },P_{\pm \beta }\}$ all satisfy $%
P^{\dagger }=P$ and are mutually commuting\cite{Chisholm}. Consequently, in
adopting Eq.\ (\ref{current}) as the form for the particle current, the only
surviving terms are those arising from within the same column. The remaining
columns $\{2,3,5,8\}$ have algebraic forms akin to those of table \ref
{table1}, but are $P_{-3}$ projections and therefore do not contribute. The
total current obtained from simply adding the algebraic equivalents of all
eight column-spinor doublets into a single element $\Psi $ is then
\begin{eqnarray}
J &=&\Psi P_{+3}\Psi ^{\dagger }  \nonumber \\
&=&\sum_{P_{+3}}\sum_{u,l}\lceil J_{(i)}^{\mu }\rfloor e_{\mu }+\text{%
(higher-dim. terms).}  \label{current2}
\end{eqnarray}
The sum here runs over the spinors assigned to the upper and lower portions
of the $P_{+3}$ columns. This is most easily verified by computing the trace
of the matrix representations of the algebriac products $Je_{\mu }$. The
eight currents of a single generation of fermions are now incorporated into
one expression. The residual part of the current involves cross-current
terms and mass-like terms of the form $\lceil \bar{\psi}\psi \rfloor $
between the upper and lower fermions of the same column, all projected onto
higher-dimensional elements. For the moment, we will be concerned with only
the physical components of the current, deferring the higher-dimensional
results to a later discussion of the gauge coupling terms and Higgs field in
the Lagrangian.

The above construction of the spinors can also be performed without any
matrix representation scaffolding. It is sufficient to specify that the
spinor doublets should be minimal left-ideals of $C\!\ell _{7}.$ However,
the matrix representation will prove a useful instructive device when making
introductory comparisons with the conventional gauge transformations.

\section{Gauge Symmetries}

The various gauge symmetries of the standard model will now be constructed
by considering rotational transformations on the total spinor which leave
the physical parts of the current in Eq.\ (\ref{current2}) invariant. {\it %
Internal\/} rotations are transformations
\begin{equation}
\Psi \rightarrow \Psi R
\end{equation}
from the right-hand side of the spinor that leave $e_{3}$ invariant, since
then
\begin{equation}
J\rightarrow \Psi RP_{+3}R^{\dagger }\Psi ^{\dagger }=J.
\end{equation}
The space of available internal planes of rotation in $C\!\ell _{7}$ is
spanned by the set of fifteen bivector generators
\begin{equation}
e_{j}e_{k}:(j,k)\in \{1,2,4,5,6,7\},\;j<k.
\end{equation}
{\it External\/} rotations are comprised of the more limited set of
transformations
\begin{equation}
\Psi \rightarrow R^{\prime }\Psi
\end{equation}
from the left of the spinor that leave the physical vectors of the current
invariant. The external rotation generators are spanned by the six planes
\begin{equation}
e_{j}e_{k}:(j,k)\in \{4,5,6,7\},\;j<k.
\end{equation}
An additional restriction is made in that one of the spinors, namely the
right-chiral neutrino, is to be exempt from any such transformations. This
will prove sufficient to completely specify the conventional gauge structure
of the standard model.

We begin by arbitrarily assigning the lepton doublet to the first column,
with the neutrino and electron occupying the upper and lower spinors
respectively. The remaining three $P_{+3}$ columns are likewise assigned to
up and down quark doublets. In excluding the right-chiral neutrino, the six
external generators are reduced to three, since the set may be regrouped
into the three listed in table \ref{table2}, which implicitly contain the
left-chiral projector, while the right-chiral partners are discarded. Each
generator characterizes a simultaneous independent rotation about two
commuting planes. These generators satisfy
\begin{equation}
\lbrack T_{a}^{\prime },T_{b}^{\prime }]=f_{abc}T_{c}^{\prime },
\label{structure2}
\end{equation}
with the antisymmetric structure constants $f_{123}=1.$ The conventional
presence of the unit imaginary in front of $T_{c}^{\prime }$ in Eq.\ (\ref
{structure2}) has been absorbed into the properties of bivectors.

In the matrix representation (\ref{matrices}), the generators $T_{a}^{\prime
}$ are nonzero only for the central $4\times 4$ submatrix. These act on the
left-chiral portions of the spinors yielding
\begin{equation}
T_{a}^{\prime }\sim \lceil -i\sigma _{a}/2\otimes 1_{2\times 2}\rfloor _{L},
\end{equation}
where $\sigma _{a}$ are the standard Pauli matrices. Therefore, the effect
of the transformation
\begin{equation}
\Psi \rightarrow \exp (\theta _{a}T_{a}^{\prime })\Psi
\end{equation}
on the total spinor is identical to that of the prevailing $SU(2)$
prescriptions
\begin{eqnarray}
{\nu _{e} \choose e^{-}}%
_{L} &\rightarrow &\exp (-i\theta _{a}\sigma _{a}/2)%
{\nu _{e} \choose e^{-}}%
_{L},  \nonumber \\
{u \choose d}%
_{L} &\rightarrow &\exp (-i\theta _{a}\sigma _{a}/2)%
{u \choose d}%
_{L}.
\end{eqnarray}

In considering the broader set of internal transformations, it is useful to
think in terms of shuffling entire columns about in the matrix
representation, which is the only structural change a rotation solely from
the right-hand side can accomplish, aside from introducing phase factors. To
remove the right-chiral neutrino from this process we insulate the first
column by discarding any candidate generators $T$ for which $P_{1}T\neq 0$.
This determines the surviving terms $T_{1}$ through $T_{7}$ arranged in
table \ref{table3}. For example, $P_{1}T_{1}=0$ while $%
P_{1}(e_{4}e_{6}+e_{5}e_{7})\neq 0$. This reduces the degrees of freedom
from fifteen to eight, and the remaining plane $e_{1}e_{2}$ can then be
fitted into $T_{8}$ under the same restriction to complete an $SU(3)$
symmetry. These generators satisfy
\begin{equation}
\lbrack T_{a},T_{b}]=-f_{abc}T_{c},  \label{structure3}
\end{equation}
where the antisymmetric structure constants are given by
\begin{eqnarray}
f_{123} &=&1,  \nonumber \\
f_{147} &=&f_{165}=f_{246}=f_{257}=f_{345}=f_{376}=\frac{1}{2},  \nonumber \\
f_{458} &=&f_{678}=\frac{\sqrt{3}}{2}.
\end{eqnarray}
Note that a similar symmetry would have existed from the left-hand side, if
it were not that the vectors $e_{1}$ and $e_{2}$ were required to remain
invariant as elements of the spatial current. The negative sign in Eq.\ (\ref
{structure3}) has been introduced simply to maintain consistent $f_{123}$
constants shared by the two non-Abelian symmetries while avoiding subsequent
sign disparities in the gauge-coupling Lagrangian terms, which are an
artifact of the double-sided transformations.

To verify that these generators induce the same effect as in the
conventional arrangement, label columns 4, 6, and 7 as red, green and blue
respectively, then transcribe the generators $T_{a}$ back into matrix form
using (\ref{matrices}). Extracting the $4\times 4$ submatrix formed by the
rows and columns 1, 4, 6 and 7 yields only a lower-right $3\times 3$
submatrix $-i\lambda _{a}^{*}$, where $\lambda _{a}$ are the well-known
Gell-Mann matrices. Therefore, the transformation
\begin{equation}
\Psi \rightarrow \Psi \exp (\theta _{a}T_{a})
\end{equation}
leaves the first column invariant and is identical in its effect on the
remaining $P_{+3}$ spinor components to
\begin{equation}
\left(
\begin{array}{ccc}
\psi _{R}, & \psi _{G}, & \psi _{B}
\end{array}
\right) \rightarrow \left(
\begin{array}{ccc}
\psi _{R}, & \psi _{G}, & \psi _{B}
\end{array}
\right) \exp (-i\theta _{a}\lambda _{a}^{*}/2),
\end{equation}
which is equivalent to the more familiar
\begin{equation}
\left(
\begin{array}{l}
\psi _{R} \\
\psi _{G} \\
\psi _{B}
\end{array}
\right) \rightarrow \exp (-i\theta _{a}\lambda _{a}/2)\left(
\begin{array}{l}
\psi _{R} \\
\psi _{G} \\
\psi _{B}
\end{array}
\right) .
\end{equation}
The difference here is that the symmetry arises naturally from the geometric
architecture and is not formed by arbitrarily imposing an $SU(3)$ symmetry
acting in some abstract space. Of course, one may deduce the effects of both
of these sets of generators by dealing strictly with the algebraic elements
and reach the same conclusions. It is a useful exercise for some of the
manipulations that will follow.

There remains one symmetry which has not been exploited as yet. We may
consider a synchronized double-sided rotation which has the effect of
various phase transformations on the spinors and which conspires to cancel
out in the case of the right-chiral neutrino. The sole candidate for such an
operation is
\begin{equation}
\Psi \rightarrow \exp (\theta _{0}T_{0}^{\prime })\Psi \exp (\theta
_{0}T_{0}),
\end{equation}
with the $U(1)$ double-sided generators given by
\begin{eqnarray}
T_{0}^{\prime } &=&\beta _{1}e_{4}e_{5}+\beta _{2}e_{6}e_{7},  \nonumber \\
T_{0} &=&\beta _{3}e_{1}e_{2}+\beta _{4}e_{4}e_{5}+\beta _{5}e_{6}e_{7},
\end{eqnarray}
where the $\beta _{k}$ are real coefficients satisfying
\begin{equation}
\beta _{1}-\beta _{2}+\beta _{3}+\beta _{4}-\beta _{5}=0.
\end{equation}
The latter restriction follows directly from the projector structure of the
right-chiral neutrino spinor. As this is to represent a distinct symmetry,
the left- and right-side generators must commute with all $SU(2)$ and $SU(3)$
generators respectively. This imposes the additional constraints
\begin{eqnarray}
\beta _{1} &=&-\beta _{2},  \nonumber \\
\beta _{3} &=&\beta _{4}=-\beta _{5}.  \label{constraints}
\end{eqnarray}
The trivial solution may be normalized to
\begin{eqnarray}
T_{0}^{\prime } &=&\frac{1}{2}(e_{5}e_{4}+e_{6}e_{7}),  \nonumber \\
T_{0} &=&\frac{1}{3}(e_{1}e_{2}+e_{4}e_{5}+e_{7}e_{6}).  \label{u1}
\end{eqnarray}
Applying this operation to each spinor in turn proves to be identical to the
conventional $U(1)_{Y}$ transformation
\begin{equation}
\lceil \psi _{(j)}\rightarrow \exp (-i\theta _{0}Y_{(j)})\psi _{(j)}\rfloor ,
\end{equation}
with the $Y_{j}$ weak hypercharge assignments listed in table \ref{table4}.
This is most easily verified by examining the matrix representation of the
infinitesimal transformation
\begin{equation}
\Psi \rightarrow \Psi +\theta _{0}T_{0}^{\prime }\Psi +\Psi \theta _{0}T_{0}
\end{equation}
for each of the $P_{+3}$ columns. All of these weak hypercharge values now
arise naturally from a single algebraic operator which was derived from
symmetry constraints, and there is no need to artificially insert the
charges for each particle.

In order to derive the electromagnetic charges we must look for linear
combinations of the generators $\{T_{0}^{\prime }\oplus T_{0},T_{3}^{\prime
}\}$ yielding assignments that are invariant under the parity transformation
of Eq.\ (\ref{parity}). Note that the generators $T_{3}$ and $T_{8}$ have
been excluded here since no linear combination has a uniform effect on the
three quarks. Since the generators from the left must commute with the
parity operator $e_{1}e_{2}e_{3}e_{4}$, the only possibility is to isolate
the generator $e_{6}e_{7}=(T_{3}^{\prime }+\frac{1}{2}T_{0}^{\prime })$ on
the left. The $U(1)_{em}$ symmetry is then specified by the generators
\begin{eqnarray}
T_{em}^{\prime } &=&\frac{1}{2}e_{6}e_{7},  \nonumber \\
T_{em} &=&\frac{1}{6}(e_{1}e_{2}+e_{4}e_{5}+e_{7}e_{6})
\end{eqnarray}
operating simultaneously from the left and right respectively. It is readily
verified that these furnish the correct electromagnetic charge assignments
for each of the fermions with
\begin{equation}
\Psi \rightarrow \exp (\theta _{em}T_{em}^{\prime })\Psi \exp (\theta
_{em}T_{em})
\end{equation}
being identical to
\begin{equation}
\lceil \psi _{(j)}\rightarrow \exp (-i\theta _{em}Q_{(j)})\psi _{(j)}\rfloor
.
\end{equation}
This is essentially the basis of the Gell-Mann-Nishijima formula\cite
{GMN1,GMN2}
\begin{equation}
Q=T_{3}+\frac{1}{2}Y.  \label{GMN}
\end{equation}

One may be tempted here to think that the restrictions in Eq.\ (\ref
{constraints}) are simply a circuitous way of establishing the same
situation, since Eq.\ (\ref{GMN}) is contrived to satisfy the symmetry
constraints
\begin{eqnarray}
Y(\nu _{L}) &=&Y(e_{L}),\quad Y(u_{L})=Y(d_{L}),  \nonumber \\
Q(\nu _{L}) &=&Q(\nu _{R}),\quad Q(u_{L})=Q(u_{R}),  \nonumber \\
Q(e_{L}) &=&Q(e_{R}),\quad Q(d_{L})=Q(d_{R}),
\end{eqnarray}
which completely determines the relative lepton assignments if $Y(\nu
_{R})=0 $ is imposed. However, no analogy of the latter restriction exists
in the quark sector, as there are no such assumptions concerning $Y(u_{R})$.
This leaves the quark system indeterminate with an infinite choice of
possible charges under Eq.\ (\ref{GMN}). The solution in Eq.\ (\ref{u1})
contains additional structural information which is lost in the standard
prescription and effectively {\em derives the observed charge assignments
for the quarks}. Ultimately, this may be traced to the identification of the
privileged subspace of bivectors within the larger algebra, a choice which
is unavailable in the conventional notation.

All of the above transformations may now be combined into a single
expression
\begin{equation}
\Psi \rightarrow \exp (\theta _{0}T_{0}^{\prime }+\theta _{a}^{\prime
}T_{a}^{\prime })\Psi \exp (\theta _{0}T_{0}+\theta _{b}T_{b}),
\end{equation}
which exhausts the plane-rotational symmetries under the condition that the
right-chiral neutrino is to be disengaged from all gauge transformations.

\section{Lagrangian Terms}

In this section, the preceding gauge formalism and overall strategy of
consolidating terms will be applied to the Lagrangian density. The previous
disregard for higher-dimensional products in the current will now be
justified in that the form chosen for the fermion terms extracts only select
components of the current.

Consider the algebraic form
\begin{equation}
\Psi ^{\dagger }\bar{e}_{\mu }\Psi .  \label{components}
\end{equation}
Since the matrix representations for the vectors of $C\!\ell _{7}$ are
Hermitian, the matrix $M$ associated with any general element $K\in C\!\ell
_{7}$ satisfies the isomorphism
\begin{equation}
(M^{*})^{T}\sim K^{\dagger }.
\end{equation}
The term $\Psi ^{\dagger }$ may then be viewed as a stack of complex
conjugate row spinors. As multiplication by $\bar{e}_{\mu }$ from the left
cannot mix the columns of $\Psi $, the diagonal of the matrix representation
of (\ref{components}) contains the contraction of each row spinor with
elements from its associated column spinor. From the representation of $%
e_{\mu }$ in (\ref{matrices}), it is then clear that

\begin{equation}
\langle \Psi ^{\dagger }\bar{e}_{\mu }\Psi \rangle _{s}=\sum_{f}\lceil
J_{\mu }^{(f)}\rfloor ,
\end{equation}
where the s-bracket denotes the real scalar part. Note that the sum includes
the lower chiral-inverted spinor, since the sign of $\gamma ^{0}\gamma ^{k}$
is reversed in the lower right $4\times 4$ submatrix of $e_{k}$. Taking the
real scalar part here, and with later terms, is simply convenient shorthand
notation for what could also be achieved by adding various symmetric
counterparts to the algebraic expressions.

It is sometimes convenient to invoke the algebraic equivalent of the
square-matrix trace theorem
\begin{equation}
tr(AB)=tr(BA)
\end{equation}
when dealing with the real scalar part. For example, we have
\begin{equation}
\langle \Psi ^{\dagger }\bar{e}_{\mu }\Psi \rangle _{s}=\langle \bar{e}_{\mu
}\Psi \Psi ^{\dagger }\rangle _{s},
\end{equation}
which is then seen to be manifestly gauge invariant for physical vectors $%
e_{\mu }$ since the physical part of the current $\Psi \Psi ^{\dagger }$
remains invariant.

The algebraic spatial derivative is defined as
\begin{equation}
\bar{\partial}\equiv \partial _{0}+\partial _{1}e_{1}+\partial
_{2}e_{2}+\partial _{3}e_{3}
\end{equation}
operating to the right. A unidirectional derivative is sufficient since only
the real scalar part is to be extracted, symmetrizing the operation
regardless. In analogy to the usual fermion derivative term in the
Lagrangian, we then have
\begin{equation}
{\cal L}_{\partial }=\langle \Psi ^{\dagger }i\bar{\partial}\Psi \rangle
_{s}=\lceil \sum_{f}\bar{\psi}_{(f)}i\gamma ^{\mu }\partial _{\mu }\psi
_{(f)}\rfloor .  \label{Ld}
\end{equation}
It must be emphasized that although this expression appears similar to the
conventional version, it contains all fermion fields of one generation in a
single algebraic term.

Explicitly writing $i=-\bar{e}_{0}e_{1}\bar{e}_{2}e_{3}\bar{e}_{4}e_{5}\bar{e%
}_{6}e_{7},$ the Lagrangian component (\ref{Ld}), for example, is then
manifestly Lorentz invariant via
\begin{equation}
{\cal L}_{\partial }\rightarrow \langle \Psi ^{\dagger }L^{\dagger }\bar{L}%
^{\dagger }iL^{\dagger }\bar{L}^{\dagger }\bar{\partial}\bar{L}L\Psi \rangle
_{s}={\cal L}_{\partial }.
\end{equation}

To ensure local gauge invariance, we introduce gauge fields $%
\{B,W_{a},G_{a}\}\in $ $C\!\ell _{3}$ which transform according to
\begin{eqnarray}
\bar{B} &\rightarrow &\bar{B}+\frac{2}{g^{\prime }}\bar{\partial}\theta _{0},
\nonumber \\
\bar{W}_{a} &\rightarrow &\bar{W}_{a}+\frac{1}{g}\bar{\partial}\theta
_{a}^{\prime }+f_{abc}\theta _{b}^{\prime }\bar{W}_{c},  \nonumber \\
\bar{G}_{a} &\rightarrow &\bar{G}_{a}+\frac{1}{g_{s}}\bar{\partial}\theta
_{a}+f_{abc}\theta _{b}\bar{G}_{c}  \label{Gtrans}
\end{eqnarray}
into the additional Lagrangian terms
\begin{eqnarray}
{\cal L}_{1} &=&-\frac{g^{\prime }}{2}\langle \Psi ^{\dagger }i\bar{B}%
(T_{0}^{\prime }\Psi +\Psi T_{0})\rangle _{s},  \nonumber \\
{\cal L}_{2} &=&-g\langle \Psi ^{\dagger }i\bar{W}_{a}T_{a}^{\prime }\Psi
\rangle _{s},  \nonumber \\
{\cal L}_{3} &=&-g_{s}\langle \Psi ^{\dagger }i\bar{G}_{a}\Psi T_{a}\rangle
_{s}.  \label{L3}
\end{eqnarray}
It is sufficient to verify invariance to first order in $\theta $ using the
infinitesimal transformation
\begin{equation}
\Psi \rightarrow \Psi +(\theta _{a}^{\prime }T_{a}^{\prime }+\theta
_{0}T_{0}^{\prime })\Psi +\Psi (\theta _{0}T_{0}+\theta _{a}T_{a}).
\end{equation}
Note that $T^{\dagger }=-T$ for all generators, and all external $T^{\prime
} $ commute with the physical gauge fields.

Rather than belabor the comparison of these terms with conventional
expressions, it will simply be mentioned that the usual charge currents are
contained within Eq.\ (\ref{L3}). For example,
\begin{equation}
{\cal L}_{1}=-\frac{g^{\prime }}{2}\lceil iY^{(f)}B^{\mu }J_{\mu
}^{(f)}\rfloor .
\end{equation}
Note that $T_{0}$ by itself would be the singlet generator associated with
the often ruminated ``ninth gluon'', which has now been absorbed into the
definition of the weak hypercharge.

In constructing the free-field expressions, the design here is to
consolidate the internal and external transformations into two separate
terms. The physical part of the tensor associated with each generator $T_{c}$
occupies the six vectors and bivectors of $C\!\ell _{3}$ and may written in
the form
\begin{equation}
F_{c}=(\bar{\partial}W_{c}-\bar{W}_{c}\partial )-g\bar{W}_{a}W_{b}f_{abc}.
\end{equation}
The full generator portion is handled through the contraction
\begin{equation}
{\cal L}_{F}=-\frac{1}{2}\langle F_{a}^{\prime }F_{b}^{\prime
}\{T_{a}^{\prime },T_{b}^{\prime }\}+F_{a}F_{b}\{T_{a},T_{b}\}\rangle _{s}.
\label{Ffield}
\end{equation}
where $a=0$ and $b=0$ are now included in the sum, since $T_{0}^{\prime }$
and $T_{0}$ commute with all other generators on their respective sides and
also do not contract any scalar elements with them. The anticommutator in
Eq.\ (\ref{Ffield}) is evidently necessary in the case of internal rotations
since products such as $T_{4}T_{5}=\frac{1}{8}(e_{7}e_{6}+e_{2}e_{1})$ would
otherwise introduce spurious terms through the presence of a physical plane.
This also ensures that under the gauge transformations of Eq.\ (\ref{Gtrans}%
), which induce the transformation
\begin{equation}
F_{a}\rightarrow F_{a}+f_{abc}\theta _{b}F_{c},
\end{equation}
the scalar part in (\ref{Ffield}) remains invariant since the only purely
physical element contracted by the generators is then the identity element.

The main reason for displaying the free-field terms in this manner is to
emphasize a key point concerning the field normalizations. The $W$ and $G$
fields may be entered into Eq.\ (\ref{Ffield}) directly, since they share a
common factor of
\begin{equation}
\langle T_{W}^{2}\rangle _{s}=\langle T_{G}^{2}\rangle _{s}=-\frac{1}{8}.
\end{equation}
In the case of $B$ we have
\begin{equation}
\langle T_{0}^{\prime 2}\rangle _{s}=-\frac{1}{2},\quad \langle
T_{0}^{2}\rangle _{s}=-\frac{1}{3},
\end{equation}
and are obliged to insert $W_{0}=G_{0}=\sqrt{3/20}B$ in order to recover the
conventional expression
\begin{equation}
{\cal L}_{F}=-\frac{1}{4}\lceil B_{\mu \nu }B^{\mu \nu }+W_{\mu \nu }\cdot
W^{\mu \nu }+G_{\mu \nu }\cdot G^{\mu \nu }\rfloor .
\end{equation}
At unification energies, where one would expect pure geometry to dominate,
the gauge transformations of $W_{a}$ and $G_{a}$ via Eq.\ (\ref{Gtrans})
should share a conjoint coupling constant relating the geometric rotation
angle to the field strengths. It then follows immediately that the bare
coupling constants obey
\begin{equation}
\frac{g}{g_{s}}=1,\quad \tan \theta _{w}\equiv \frac{g^{\prime }}{g}=\sqrt{%
\frac{3}{5}},
\end{equation}
which results in a Weinberg angle\cite{Weinberg} of $\sin ^{2}\theta _{w}=%
\frac{3}{8}.$ Radiative corrections are assumed to lower the weak mixing
angle to the observed value of $\sin ^{2}\theta _{w}\simeq 0.23$ at
accelerator energies. It has not been examined as yet what role the extra
space-like dimensions might play in such a renormalization procedure carried
out within this framework.

Conspicuously absent in the previous Lagrangian terms are those projecting
out the higher-dimensional mass-like components of the current in Eq.\ (\ref
{current2}). It should be mentioned that these reside on the 8-dimensional
Lorentz-invariant space $\{e_{4},e_{5}\}\otimes P_{\pm \beta }\oplus
\{e_{6},e_{7}\}\otimes P_{\pm \alpha },$ and provide a natural inclusion of
the Higgs complex scalar isodoublet $H$ and anti-Higgs $\widetilde{H}$
through the identification
\begin{eqnarray}
H &=&-(\phi _{1}e_{6}+\phi _{2}e_{7})P_{+\alpha }-(\phi _{3}e_{5}+\phi
_{4}e_{4})P_{-\beta }  \nonumber \\
&\sim &\lceil
{\phi _{1}+i\phi _{2} \choose \phi _{3}+i\phi _{4}}%
\rfloor ,  \nonumber \\
\widetilde{H} &=&-(\xi _{1}e_{6}+\xi _{2}e_{7})P_{-\alpha }-(\xi
_{3}e_{5}+\xi _{4}e_{4})P_{+\beta }  \nonumber \\
&\sim &\lceil
{-\xi _{3}+i\xi _{4} \choose \xi _{1}-i\xi _{2}}%
\rfloor ,
\end{eqnarray}
in the expression
\begin{equation}
{\cal L}_{M}=\frac{G_{f}}{\sqrt{2}}\langle \Psi ^{\dagger }(H+\widetilde{H}%
)\Psi \rangle _{s}.  \label{Lhiggs}
\end{equation}
These algebraic elements form a carrier space for the set of external gauge
generators where, for example, the transformation required for gauge
invariance,
\begin{equation}
H\rightarrow \exp (\theta _{0}T_{0}^{\prime }+\theta _{a}^{\prime
}T_{a}^{\prime })H\exp (-\theta _{0}T_{0}^{\prime }-\theta _{b}^{\prime
}T_{b}^{\prime }),
\end{equation}
is equivalent to the conventional notation
\begin{equation}
{\phi ^{+} \choose \phi ^{0}}%
\rightarrow \exp (-iY\theta _{0}-i\theta _{a}\sigma _{a}/2)%
{\phi ^{+} \choose \phi ^{0}}%
.
\end{equation}
The weak hypercharge assignment of $Y=1$ $(Y=-1)$ for the Higgs particle
(antiparticle) is recovered naturally from the double-sided algebraic
transformation. Here again, the Higgs field is no longer an artificial
appendage cast in some abstract space, but emerges readily from the geometry
and is associated with the higher-dimensional vector components of the
current.

The form of Eq.\ (\ref{Lhiggs}) is an illustrative example for equal fermion
masses. Distinct masses may be introduced through weighted projectors on
both sides of the spinor set $\Psi $. This particular area is not understood
at present, but it is encouraging that there seems to be additional
structure to work with here.

For completeness, the remaining parts of the minimal Higgs sector may also
be written algebraically. The gauge-invariant free-field and potential terms
respectively are
\begin{eqnarray}
{\cal L}_{H} &=&\langle (\bar{\partial}H-\frac{g^{\prime }}{2}\bar{B}%
[T_{0}^{\prime },H]-g\bar{W}_{j}[T_{j}^{\prime },H])^{2}\rangle _{s},
\nonumber \\
{\cal L}_{V} &=&-\langle \mu H^{2}+\lambda H^{4}+\cdots \,\rangle _{s},
\end{eqnarray}
where Eq.\ (\ref{hi-com}) provides for the Minkowski contraction of the
physical components. The gauge symmetry may broken by choosing a vacuum
expectation value
\begin{equation}
H_{0}=-\upsilon e_{5}P_{-\beta },\;\upsilon \equiv -\mu ^{2}/\lambda ,
\end{equation}
analogous to the conventional choice. This leads directly to the
vector-boson mass relations of the Weinberg-Salam model\cite{Weinberg,Salam}%
, the relevent initial term being
\begin{equation}
{\cal L}_{H_{0}}=\frac{\upsilon ^{2}}{8}[g^{2}\bar{W}_{j}W_{j}-gg^{\prime }(%
\bar{W}_{3}B+\bar{B}W_{3})+g^{\prime 2}\bar{B}B]+\cdots .
\end{equation}

\section{Conclusion}

A common criticism on the application of Clifford algebras is that, despite
any alluring framework that may be constructed, geometric approaches are
merely reformulations of conventional expressions and are consequently
devoid of additional insights. In the present work, such an amelioration
prompted the addition of extra space-like dimensions and was essential in
identifying the otherwise obscured geometric subspace in which the gauge
symmetries abide. These symmetries are no longer relegated to abstract
spaces, but are seen to arise naturally from the architecture of
higher-dimensional spinors. As a consequence, the proposed unification of
the elementary-particle forces follows from purely geometric concerns.

Note that the value of $\sin ^{2}\theta _{w}=\frac{3}{8}$ at unification
energies is identical to the often touted result from minimal $SU(5)$ grand
unification\cite{su(5)}. This is not surprising, as both originate from the
normalization of the weak hypercharge operator. However, the contention here
is distinctly different. In the geometric approach, the second
`symmetry-breaking' of the strong and electroweak forces arises naturally by
invoking double-sided transformations. The introduction of a master group
such as $SU(5)$ to accomplish this adds extra gauge bosons and the hierarchy
problem associated with their experimental absence. Furthermore, there is
seldom any physical justification for a particular abstract master group,
other than its suiting a selection process under various constraints. This
has been done here to some extent in that the choice of $C\!\ell _{7}$ was
to accommodate the number of observed fermions. However, the notion that
higher dimensions are directly responsible for the presence of gauge groups
is more fundamental, as the physical basis for those symmetries is apparent.

An obvious deficiency of this model in its present form is the need to
suppress the right-chiral neutrino. Also lacking is a geometric rationale
for the inclusion of three generations of fermions and for some indication
as to the origin of their disjointed masses. Work is continuing on these and
other aspects of the standard model within this framework.

\acknowledgments
The author would like to thank W. E. Baylis for both supervisory support and
countless valuable discussions, without which this work would not have been
possible. This work was supported by the Natural Sciences and Engineering
Research Council of Canada.

\begin{table}[tbp]
\caption{The algebraic $P_{+3}$ spinors.}
\label{table1}
\begin{tabular}{ccc}
column \# & lower spinor & upper spinor \\
\tableline 1 & $\sqrt{8}(\psi _{R}e_{5}e_{6}+\psi _{L}e_{6}e_{1})P_{-\alpha
}P_{+\beta } $ & $\sqrt{8}(\psi _{R}+\psi _{L}e_{1}e_{5})P_{-\alpha
}P_{+\beta }$ \\
4 & $\sqrt{8}(\psi _{R}e_{1}e_{6}+\psi _{L}e_{5}e_{6})P_{+\alpha }P_{+\beta
} $ & $\sqrt{8}(\psi _{R}e_{5}e_{1}+\psi _{L})P_{+\alpha }P_{+\beta }$ \\
6 & $\sqrt{8}(\psi _{R}e_{1}e_{5}+\psi _{L})P_{-\alpha }P_{-\beta }$ & $%
\sqrt{8}(\psi _{R}e_{1}e_{6}+\psi _{L}e_{6}e_{5})P_{-\alpha }P_{-\beta }$ \\
7 & $\sqrt{8}(\psi _{R}+\psi _{L}e_{5}e_{1})P_{+\alpha }P_{-\beta }$ & $%
\sqrt{8}(\psi _{R}e_{6}e_{5}+\psi _{L}e_{6}e_{1})P_{+\alpha }P_{-\beta }$%
\end{tabular}
\end{table}

\begin{table}[tbp]
\caption{The algebraic SU(2) generators.}
\label{table2}
\begin{tabular}{l}
$T_{1}^{\prime }=\frac{1}{4}(e_{6}e_{4}+e_{5}e_{7})$ \\
$T_{2}^{\prime }=\frac{1}{4}(e_{4}e_{7}+e_{5}e_{6})$ \\
$T_{3}^{\prime }=\frac{1}{4}(e_{4}e_{5}+e_{6}e_{7})$%
\end{tabular}
\end{table}

\begin{table}[tbp]
\caption{The algebraic SU(3) generators.}
\label{table3}
\begin{tabular}{l}
$T_{1}=\frac{1}{4}(e_{6}e_{4}+e_{5}e_{7})$ \\
$T_{2}=\frac{1}{4}(e_{7}e_{4}+e_{6}e_{5})$ \\
$T_{3}=\frac{1}{4}(e_{4}e_{5}+e_{6}e_{7})$ \\
$T_{4}=\frac{1}{4}(e_{1}e_{7}+e_{2}e_{6})$ \\
$T_{5}=\frac{1}{4}(e_{6}e_{1}+e_{2}e_{7})$ \\
$T_{6}=\frac{1}{4}(e_{1}e_{4}+e_{2}e_{5})$ \\
$T_{7}=\frac{1}{4}(e_{5}e_{1}+e_{2}e_{4})$ \\
$T_{8}=\frac{1}{4\sqrt{3}}(2e_{1}e_{2}+e_{5}e_{4}+e_{6}e_{7})$%
\end{tabular}
\end{table}

\begin{table}[tbp]
\caption{Charge assignments for fermions.}
\label{table4}
\begin{tabular}{cccc}
& $T_{3}$ & $Y$ & $Q$ \\
\tableline $\nu _{L}$ & 1/2 & -1 & 0 \\
$\nu _{R}$ & 0 & 0 & 0 \\
$e_{L}$ & -1/2 & -1 & -1 \\
$e_{R}$ & 0 & -2 & -1 \\
$u_{L}$ & 1/2 & 1/3 & 2/3 \\
$u_{R}$ & 0 & 4/3 & 2/3 \\
$d_{L}$ & -1/2 & 1/3 & -1/3 \\
$d_{R}$ & 0 & -2/3 & -1/3
\end{tabular}
\end{table}

\end{document}